\documentclass[aip,rsi,reprint,graphicx]{revtex4-1} 

\usepackage[pdftex]{hyperref,color,graphicx}
\usepackage[utf8]{inputenc}
\hypersetup{colorlinks=true,linkcolor=blue,citecolor=blue,filecolor=blue,urlcolor=blue}
\usepackage{amsfonts,amssymb,amsmath,siunitx}
\usepackage[english]{babel}
\usepackage{tabularx}
\usepackage[top=1.5cm,left=1.5cm,right=1.6cm,bottom=1.6cm]{geometry} 

\newif\ifusebibfile
\usebibfiletrue   
\usebibfilefalse

\newcommand{\figref}[2]{\hyperref[#1]{\ref{#1}(#2)}}

\begin{document}
\selectlanguage{english}

\title{Note: Ultra-low birefringence dodecagonal vacuum glass cell}

\author{Stefan Brakhane}
\email{brakhane@iap.uni-bonn.de}
\author{Wolfgang Alt}
\author{Dieter Meschede}
\author{Carsten Robens}
\author{Geol Moon}
\author{Andrea Alberti} 	
\affiliation{Institut f\"ur Angewandte Physik, Universit\"at Bonn,
Wegelerstr.~8, D-53115 Bonn, Germany}
\date{\today}


\begin{abstract}
We report on an ultra-low birefringence dodecagonal glass cell for ultra-high vacuum applications. The epoxy-bonded trapezoidal windows of the cell are made of SF57 glass, which exhibits a very low stress-induced birefringence.
We characterize the birefringence $\Delta n$ of each window with the cell under vacuum conditions, obtaining values around $\num{e-8}$.
After baking the cell at $\SI{150}{\degreeCelsius}$, we reach a pressure below $\SI{e-10}{\milli \bar}$. In addition, each window is antireflection coated on both sides, which is highly desirable for quantum optics experiments and precision measurements.
\end{abstract}

\maketitle
Modern experiments for the investigation of cold atom ensembles require an ultra-high vacuum apparatus with two main characteristics: (1) very large optical access \cite{Soltan-Panahi:2011} and (2) accurate  preservation of laser beam properties such as the state of polarisation and the wavefront quality \cite{bloch,steffenPNAS}.
The standard approach consists in using a metal vacuum chamber in combination with viewports.
To minimize the birefringence caused by mechanical stress in the viewport, special procedures have been developed to mount the windows \cite{windowMounting,lowRetardanceWindow,windowOrientation}, and glass materials with extremely low stress-optical coefficients have been employed \cite{pressureCell}.

As an alternative to metal chambers, vacuum glass cells are widely used since they generally exhibit less birefringence than conventional vacuum viewports \cite{Steffen}.
Furthermore, owing to their small volume, glass cells can be combined with compact electromagnetic coils, which allows one to generate strong magnetic fields \cite{feshbach} and field gradients \cite{Sherson:2010} that can be switched on a short time scale.
Glass cells are commonly produced by diffusion bonding of the individual glass windows.
The high temperatures involved in this bonding process, however, limit the application of optical coatings to the outside of the cell after its assembly.
Reflections by the inner surfaces can constitute a severe shortcoming since they yield stray light and undesired optical standing waves.

Only recently, glass cells bonded by optical contact have become commercially available with double-sided antireflection coating and in more versatile geometries \cite{OpticalContactCompanies}.
This bonding method, on the other hand, requires the contact surfaces to be polished to highest precision and the different components to be aligned with very high mechanical accuracy.
For that reason, optical contact has been so far applied to vacuum cells with up to eight facets, and furthermore using only standard glass materials.

In this paper, we report on an epoxy-bonded and double-side antireflection-coated vacuum glass cell, which combines the excellent optical access of the dodecagonal geometry with the exceptionally low birefringence of Schott SF57 glass \cite{pressureCell}.
This material features one of the lowest available stress-optical coefficients, which is about two orders of magnitude below that of conventional glass materials.
The application of this glass allows us to construct vacuum cells with ultra-low birefringence ($\Delta n/n<10^{-7}$).
Furthermore, lead glasses like SF57 are particularly suited for vacuum applications due to their low permeability to Hydrogen and Helium \cite{vacuumHandbook}, but care must be taken owing to their high sensitivity to temperature changes and mechanical shocks.


\begin{figure}[t]
	\includegraphics[width = 85mm]{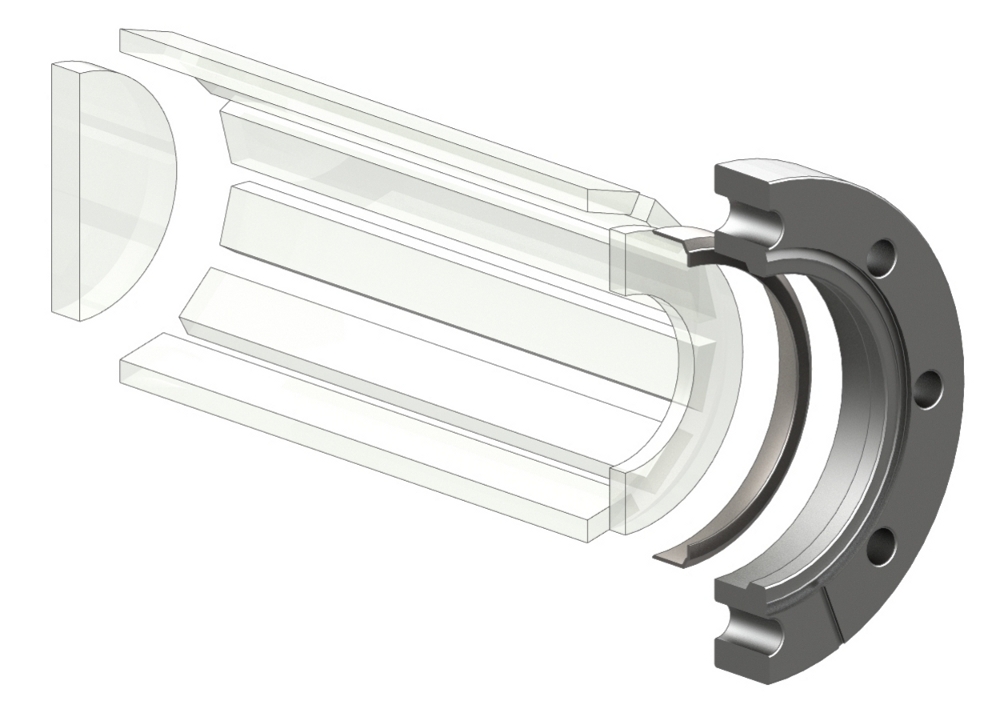}
	\caption{\label{fig1}Exploded view of half of the vacuum glass cell consisting of: a round cover glass (thickness $\SI{9}{\milli\metre}$, outer diameter $\SI{56}{\milli\metre}$), 12 windows with a trapezoidal cross section (inner aperture size $\SI{13}{\milli\metre} \times \SI{150}{\milli\metre}$, thickness $\SI{5}{\milli\metre}$), a glass ring (thickness $\SI{5}{\milli\metre}$, inner diameter $\SI{40}{\milli\metre}$, outer diameter $\SI{68}{\milli\metre}$), and a DN63CF stainless steel (316L) flange featuring a Tantalum weld ring. The inner diameter is around $\SI{48.5}{\milli\meter}$.
}
\end{figure}

The cell consists of twelve windows with a trapezoidal cross section forming a cylinder-like structure, shown in Fig.~\ref{fig1}.
The structure is closed on one side by a round cover glass and is connected on the other side to a glass ring of the same material.
All glass components are polished with surface flatness $\lambda/20$ ($\lambda=\SI{866}{\nano\meter}$) and  antireflection coated on both sides except for the ring.
Moreover, the cell exhibits a compact geometry and yet allows one to house within it scientific components, e.g., a microscope objective for high-resolution imaging.

To bond the cell windows, different thermally-cured epoxy adhesives that fulfill the NASA low outgassing standard (ASTM E595) have been compared.
We find that the EpoTek H77 adhesive containing filling particles yields lower stress-induced birefringence  compared to unfilled epoxy adhesives (EpoTek 353ND, EpoTek 353T).
Moreover, the filling particles ensure a minimum thickness of the adhesive itself, which is advantageous for handling.
We make use of an automated adhesive dispenser to plot a reproducible line of epoxy resin at the contact surfaces, which yields a homogeneous epoxy layer when two windows are contacted.
Surface roughness and small deviations from the ideal geometry are compensated by the glue volume, therefore allowing for relatively large tolerances in the glass cutting process.
A very slow cooling process from the maximum curing temperature of $\SI{150}{\degreeCelsius}$ down to room temperature over several hours leads to a further reduction of the amount of stress-induced birefringence.

We demonstrate two different mounting procedures of the cell to a standard ConFlat flange: (1) The glass structure (coefficient of thermal expansion (CTE) $\SI{8.6e-6}{\per\kelvin}$\;\cite{catalogueSchott}) is glued to a commercially available non-magnetic stainless steel viewport flange equipped with a Tantalum weld ring (CTE $\SI{6.5e-6}{\per\kelvin}$\;\cite{thermalExpansion}), as illustrated in Fig.~\ref{fig1}.
The thin and relatively soft weld ring prevents the formation of critical stress levels within the glass material when the flange is deformed, e.g.\ during tightening of the bolts or temperature changes.
(2) Alternatively, a second glass cell has been directly glued to a bored Titanium blind flange (CTE $\SI{8.4e-6}{\per\kelvin}$\;\cite{thermalExpansion,knifeEdgeSeal}) with hole diameter of $\SI{40}{\milli\metre}$. Unless otherwise stated, we hereafter refer to the cell bonded to the weld ring.

\begin{figure}[t]
	\includegraphics[width = 85mm]{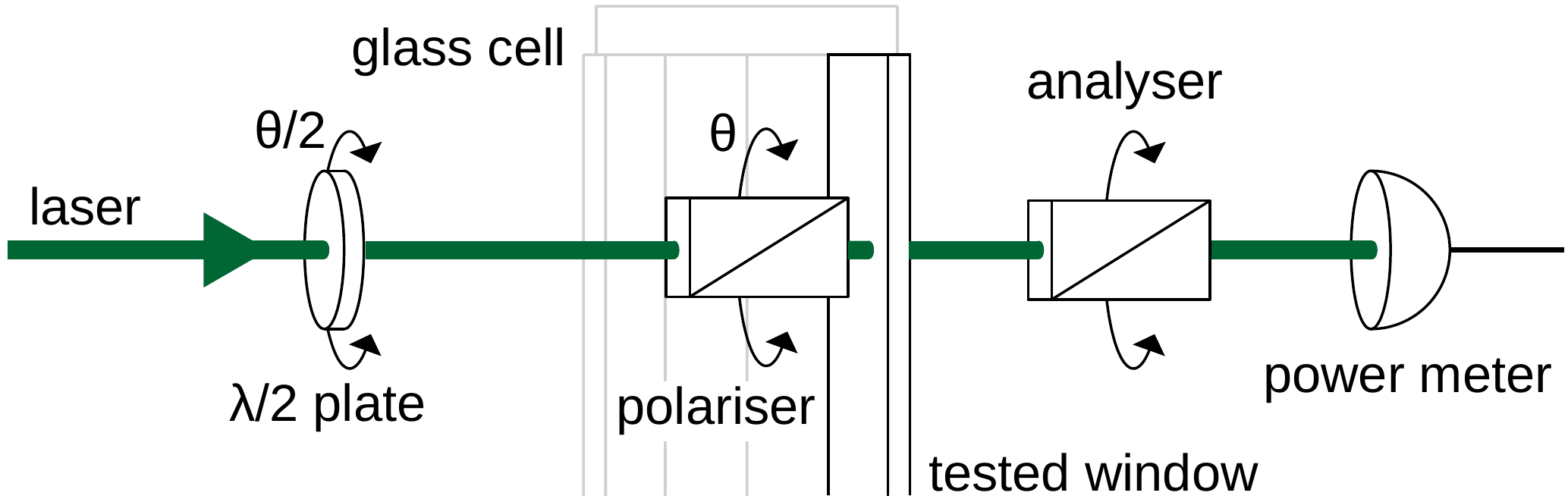}
	\caption{\label{fig2}%
	A linearly polarised probe laser beam crosses a window. The linear polarisation angle $\theta$ is set by the Glan-laser polariser (Thorlabs Inc.) inside the glass cell.
A power meter measures the minimum and maximum intensity after a second rotating Glan-laser polariser, which is used as an analyser.
To suppress stray light, the power meter is placed at about $\SI{2}{\meter}$ distance.}\vspace{-2.5mm}
\end{figure}

The temperature during the bake-out procedure is kept below $\SI{150}{\degreeCelsius}$ to avoid strain from the unmatched thermal expansion of glass and metal, although the epoxy adhesive itself withstands temperatures up to $\SI{260}{\degreeCelsius}$.
After the bake-out procedure, we achieve a pressure around $\SI{2e-10}{\milli \bar}$ in a vacuum system comprising a ion getter pump (Agilent VacIon Plus 55) and a non-evaporative getter pump (SAES Getters CapaciTorr D400-2) in addition to the vacuum cell.
The pressure has been monitored for more than six months and is mainly limited by water diffusion through the epoxy adhesive \cite{vacuumHandbook}.
The pressure is subsequently reduced below $\SI{e-10}{\milli\bar}$ (the pressure limit of our vacuum gauge) by introducing Caesium vapour into the vacuum apparatus, presumably because of the strong gettering of water molecules by Caesium atoms.
A pressure in this range implies storage times of atoms in an optical dipole trap between tens of seconds and minutes \cite{scatterRates}.
In addition, we find an Helium leak rate below $\SI{8e-11}{\milli\bar.\litre\per\second}$ by performing an integral Helium leak test of the glass cell bonded to the Titanium flange. 


\begin{figure}[t]
	\includegraphics[width = 85mm]{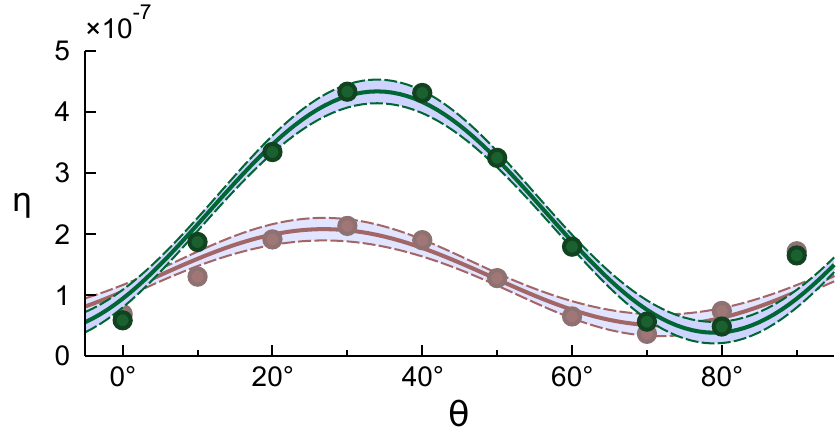}
	\caption{\label{fig3}Measured extinction ratio $\eta$ as a function of the polariser angle $\theta$ under atmospheric pressure (red) and vacuum (green). Data points refer to window no.\ 11, which exhibits the highest amount of birefringence (see Table \ref{tab1}). Solid lines represent a fit of our model to the data. Instrumental uncertainty is below the marker's size. The vertical offset is caused by stray light from the $\SI{30}{\milli\watt}$ incoming probe laser beam. Dashed lines indicate the \SI{68}{\percent}-confidence region.}\vspace{-1mm}
\end{figure}

We determine the birefringence of the evacuated cell by measuring the polarisation distortion of a linearly polarised probe laser beam crossing a single cell window.
The laser beam probes a circular region of \SI{2}{\milli\meter} diameter situated about $\SI{20}{\milli\metre}$ above the flange.
The measurement procedure is illustrated in Fig.~\ref{fig2}:
For each choice of the polarisation angle $\theta$ inside the cell, we record the maximum $I_+(\theta)$ and minimum $I_-(\theta)$ laser intensity after a rotating polarisation analyser positioned behind the cell.
The measured extinction ratio, which we define as $\eta(\theta)\equiv I_-(\theta)/I_+(\theta)$, exhibits a sinusoidal variation as a function of $\theta$, as shown in Fig.~\ref{fig3}.
The amplitude of the recorded signal allows us to determine the amount of birefringence.

We use Jones' calculus to model the transformation of polarisation by the vacuum cell.
We assume a transformation matrix $M$ of the most general form\cite{alberti:2014}
\begin{equation*}
	M = R(\beta) \cdot R(\theta_{0}) \cdot
	\begin{pmatrix}
	e^{i \phi/2} & 0\\
	0 & e^{-i \phi/2}
	\end{pmatrix}
	\cdot R(-\theta_{0}) \,,
\end{equation*}
where $\phi$ is the phase retardation, $\theta_{0}$ is the angle denoting the orientation of the optical axes parallel to the window's surface, and $\beta$ is the angle characterizing the optical activity. $R(\gamma)$ is a $2{\times}2$ rotation matrix by an angle $\gamma$.
We obtain for the intensities $I_+(\theta)$ and $I_-(\theta)$ of the setup in Fig.~\ref{fig2}:
\begin{equation}
	\label{eq:model}
I_\pm\hspace{-2pt}=\hspace{-1pt}\frac{1}{2}\pm\frac{1}{2}\sqrt{1-\sin ^2(\phi ) \sin ^2\big(2 (\theta-\theta_{0})\big)}
\end{equation}
which are independent of $\beta$.
The ratio of the two intensities yields, according to the definition above, $\eta(\theta)$, which is fitted to the experimental data shown in Fig.~\ref{fig3}, up to an offset caused by the background stray light.
The fitting procedure allows us to determine $\theta_{0}$ and $\phi$, where $\phi= \hspace{1pt} k \hspace{1pt} L \hspace{1pt} \Delta n$ ($L$ is the thickness of each glass window, $k = 2\pi/\lambda$ is the wave vector of the probe laser beam, and $\Delta n$ is the amount of birefringence).
In addition, we note that $\Delta n$ is directly related to the peak-to-peak amplitude $A$ of the signal $\eta(\theta)$ according to the formula
$$A=\frac{1-\left| \cos (k L \Delta n )\right|}{1+\left|\cos (k L \Delta n )\right|} \ \text{.}$$
For small birefringence, we have $\Delta n\approx 2\sqrt{A}/(k L)$, which shows, in this limit, the quadratic dependence of the signal's ampltiude $A$ on the amount of birefringence $\Delta n$.

\newlength{\spacingCols}
\setlength{\spacingCols}{3mm}
\begin{table*}[!t]\vspace{-2mm}
\caption{\label{tab1}Birefringence $\Delta n$ and angle $\theta_{0}$ are obtained from fitting the theoretical model (Eq.~\ref{eq:model}) to the measured extinction ratios under vacuum conditions, see Fig.~\ref{fig2}. The uncertainty is on the last digit except where otherwise stated in brackets. The angle $\theta_{0}$ is given with respect to the symmetry axis of the cell.}
\vspace{1mm}\begin{tabularx}{2\columnwidth}{>{\hspace{2mm}}p{21mm}c>{\hspace{\spacingCols}}c>{\hspace{\spacingCols}}c>{\hspace{\spacingCols}}c>{\hspace{\spacingCols}}c>{\hspace{\spacingCols}}c>{\hspace{\spacingCols}}c>{\hspace{\spacingCols}}c>{\hspace{\spacingCols}}c>{\hspace{\spacingCols}}c>{\hspace{\spacingCols}}c>{\hspace{\spacingCols}}c}
\hline\hline\\[-3mm]
{No.} & 1 & 2 & 3 & 4 & 5 & 6 & 7 & 8 & 9 & 10 & 11 & 12\\[0.3mm] \hline\\[-3mm]
\textbf{$\Delta n \;(\num{e-8})$} & 3.0(4) & 1.6(4) & 2.4(2) & 1.8(4) & 4.0(6) & 0.8(4) & 1.2(2) & 3.4(4) & 2.8(4) & 1.6(2) & 7.8(2) & 1.0(4) \\[0.3mm]
\hline\\[-3mm]
\textbf{$\theta_{0}$} & \SI{-2}{\degree} & \SI{22}{\degree} & \SI{-41}{\degree} & \SI{22}{\degree} & \SI{26}{\degree} & \SI{45}{\degree} & \SI{-18}{\degree} & \SI{5}{\degree} & \SI{-7}{\degree} & \SI{43}{\degree} & \SI{33}{\degree} & \SI{1}{\degree} \\[0.1mm]
\hline \hline
\end{tabularx}\vspace{-1mm}
\end{table*}

In Tab.~\ref{tab1} we list the obtained value of $\Delta n$ and $\theta_{0}$ for each window of the glass cell bonded to the Titanium flange.
The birefringence values are on the level of $\num{e-8}$, which is well below the typical values recorded with optical glass cells ($\num{e-7}$) and viewports ($\num{e-6}$)\; under vacuum conditions \cite{lowRetardanceWindow, Steffen}.
The amount of birefringence recorded in our vacuum cell translates into tiny retardances on the level of $\lambda/\num{5000}$, which could be further compensated by tilting an Ehringhaus compensator \cite{Ehringhaus}.
We further remark that the measured birefringence values under atmospheric and vacuum conditions are on the same order or magnitude.
For some windows we observed that the birefringence value is even reduced after evacuating the cell.
By considering the distribution of $\theta_{0}$, we infer that the mechanical stress tensor is oriented for each window arbitrarily with respect to the symmetry axis of the cell.
We further infer from repeated measurements that the birefringence does not change significantly within a few degrees deviation from normal incidence or within $\pm \SI{2}{\milli\meter}$ offset from the window's center.
In addition, similar measurements for the glass cell bonded to the Tantalum weld ring yield comparable birefringence values. For this second cell, pairs of two windows facing each other are probed together instead of each window separetely.

To conclude, we have presented methods for constructing a dodecagonal vacuum glass cell with ultra-low birefringence and double-sided antireflection coating.
The extraordinarily low birefringence of our vacuum cell is essential for modern experiments ranging from generation of synthetic gauge fields \cite{Lin:2009} and artificial spin-orbit coupling \cite{Struck:2014} to quantum technologies like coherent spin-dependent transport of atoms \cite{Alberti:2014b}. Ultra-low birefringence is also crucial for atomic clock experiments \cite{Nicholson:2015} and precision measurements of electric dipole moment \cite{zhu:2013} and vacuum polarizability \cite{DellaValle:2013}.
In addition, we have demonstrated that the cell is well suited for ultra-high vacuum apparatuses in spite of the epoxy adhesive used for the glass bonding.
In general, the epoxy bonding technique is also applicable to complex geometries with different shapes, where optical contacting is difficult.
The twelve-sided geometry, in particular, allows one to realize periodic optical potentials in different Bravais classes, such as square, hexagonal, and kagome lattices \cite{opticalLattices}.
Finally, the inner volume of the glass cell is sufficiently large to host further scientific components, for instance, an objective lens for high resolution imaging \cite{Bakr:2009}, atom chips \cite{atomChipBook} or optical cavities \cite{cavities}.

\begingroup
\renewcommand{\section}[2]{}%

\vspace*{0.2cm}

\begin{acknowledgments}
We acknowledge the financial support by the NRW-Nachwuchsforschergruppe ``Quantenkontrolle auf der Nanoskala'', ERC grant DQSIM, Deutsche Forschungsgemeinschaft (grant Forschergruppe FOR635), and EU project SIQS. In addition, SB and CR acknowledge the support by the BCGS Garduate School, and CR the support by Studienstiftung des deutschen Volkes.
\end{acknowledgments}

\endgroup

\end{document}